\documentclass[pdflatex,sn-mathphys-num]{sn-jnl}

\usepackage{graphicx}%
\usepackage{multirow}%
\usepackage{amsmath,amssymb,amsfonts}%
\usepackage{amsthm}%
\usepackage{mathrsfs}%
\usepackage[title]{appendix}%
\usepackage{xcolor}%
\usepackage{textcomp}%
\usepackage{manyfoot}%
\usepackage{booktabs}%
\usepackage{algorithm}%
\usepackage{algorithmicx}%
\usepackage{algpseudocode}%
\usepackage{listings}%

\newcommand{\revision}[1]{\textcolor{blue}{#1}}

\theoremstyle{thmstyleone}%
\theoremstyle{thmstyletwo}%

\theoremstyle{thmstylethree}%

\begin{document}

\title[Article Title]{Comparative Study of Ultrasound Shape Completion and CBCT-Based AR Workflows for Spinal Needle Interventions}


\author[1,2]{\fnm{Tianyu} \sur{Song}}
\equalcont{These authors contributed equally to this work.}

\author*[1,2]{\fnm{Feng} \sur{Li}}\email{feng.li@tum.de}
\equalcont{These authors contributed equally to this work.}

\author[1]{\fnm{Felix} \sur{Pabst}} 
\equalcont{These authors contributed equally to this work.}

\author[1,2]{\fnm{Miruna-Alexandra} \sur{Gafencu}} 

\author[1,2]{\fnm{Yuan} \sur{Bi}} 

\author[1]{\fnm{Ulrich} \sur{Eck}} 

\author[1,2]{\fnm{Nassir} \sur{Navab}} 

\affil[1]{\orgdiv{Chair for Computer Aided Medical Procedures and Augmented Reality}, \orgname{Technical University of Munich}, \orgaddress{\city{Munich}, \country{Germany}}}

\affil[2]{\orgname{Munich Center for Machine Learning (MCML)}, \orgaddress{\city{Munich}, \country{Germany}}}




\abstract{
\textbf{Purpose:} This study compares two augmented reality (AR)–guided imaging workflows, one based on ultrasound shape completion and the other on cone-beam computed tomography (CBCT), for planning and executing lumbar needle interventions. The aim is to assess how imaging modality influences user performance, usability, and trust during AR-assisted spinal procedures.

\textbf{Methods:} Both imaging systems were integrated into an AR framework, enabling in situ visualization and trajectory guidance. The ultrasound-based workflow combined AR-guided robotic scanning, probabilistic shape completion, and AR visualization. The CBCT-based workflow used AR-assisted scan volume planning, CBCT acquisition, and AR visualization. A between-subject user study was conducted and evaluated in two phases: (1) planning and image acquisition, and (2) needle insertion. 

\textbf{Results:} Planning time was significantly shorter with the CBCT-based workflow, while SUS, SEQ, and NASA-TLX were comparable between modalities. In the needle insertion phase, the CBCT-based workflow yielded marginally faster insertion times, lower placement error, and better subjective ratings with higher Trust. The ultrasound-based workflow achieved adequate accuracy for facet joint insertion, but showed larger errors for lumbar puncture, where reconstructions depended more heavily on shape completion.

\textbf{Conclusion:} The findings indicate that both AR-guided imaging pipelines are viable for spinal intervention support. CBCT-based AR offers advantages in efficiency, precision, usability, and user confidence during insertion, whereas ultrasound-based AR provides adaptive, radiation-free imaging but is limited by shape completion in deeper spinal regions. These complementary characteristics motivate hybrid AR guidance that uses CBCT for global anatomical context and planning, augmented by ultrasound for adaptive intraoperative updates.
}




\maketitle

\section{Introduction}\label{sec:Introduction}

Accurate visualization of spinal anatomy is essential for image-guided interventions such as facet joint injections and lumbar punctures, where precise localization of vertebral landmarks determines procedural safety and efficacy. Conventional guidance techniques, including fluoroscopy and computed tomography (CT), provide high anatomical accuracy but expose both patients and operators to ionizing radiation and lack real-time adaptability. Cone-beam CT (CBCT) offers intraoperative 3D imaging at reduced doses, yet it remains a static modality, insensitive to motion or deformation during procedures.

To address these limitations, ultrasound (US) has emerged as a promising alternative, offering radiation-free, real-time imaging suitable for soft-tissue and bony landmark visualization. However, US-guided spine interventions remain technically demanding due to acoustic shadowing and restricted penetration through bone, which limit visibility to posterior vertebral structures. As noted by Rekatsina~\emph{et al.} ~\cite{rekatsina2025importance}, fluoroscopy, CT, and ultrasound remain the principal imaging methods for spinal interventions, each balancing image quality, safety, and workflow integration. Comparative studies ~\cite{kimura2023comparative,viderman2023ultrasound} reported that ultrasound-guided spinal injections can reduce procedural time and adverse events relative to fluoroscopy or CT, demonstrating the growing clinical viability of ultrasound-based navigation.

Despite these advances, conventional ultrasound systems provide only partial anatomical information, forcing clinicians to infer the 3D structure from limited 2D views. Earlier research works~\cite{nagpal2015multi,azampour2024anatomy} attempted to overcome this limitation through CT-to-US registration, which incorporates preoperative CT priors into intraoperative ultrasound frames. However, such approaches are error-prone, sensitive to patient positioning, and dependent on available CT data.

Recent progress in ultrasound-based shape completion has provided a more direct and data-driven solution. Gafencu~\emph{et al.}~\cite{gafencu2024shape} proposed a probabilistic deep-learning framework that learns vertebral shape priors from CT data to reconstruct complete lumbar vertebrae from partial ultrasound inputs, accurately preserving critical anatomical landmarks such as the spinous process and facet joints.  Building on this foundation, Gafencu~\emph{et al.}~\cite{gafencu2025shape} and Li~\emph{et al.}~\cite{li20253d} extended this concept toward a robotic pipeline with automated ultrasound acquisition and real-time shape completion for enhanced spinal visualization. These developments suggest that ultrasound-based 3D reconstruction could serve as a radiation-free alternative to static CT or CBCT for intraoperative spinal visualization.

In parallel, augmented reality (AR) has emerged as a powerful tool for surgical guidance, enhancing visualization and spatial understanding during medical procedures~\citep{navab2022medical}. By overlaying anatomical models or planned trajectories directly within the operative field, AR enables intuitive, in situ navigation that reduces cognitive load compared to conventional monitor-based systems~\citep{fotouhi2020development,song2022happy}. A systematic review by \revision{Liu}~\emph{et al.}~\cite{liu2022spine} reported that AR integration in spine surgery improves spatial awareness and ergonomics while reducing radiation exposure, though challenges remain in calibration, tracking accuracy, and user comfort.

Early implementations of AR primarily focused on pedicle screw placement, one of the most demanding spinal procedures. Gibby~\emph{et al.}~\cite{gibby2019head} demonstrated the feasibility of head-mounted display (HMD)–based AR navigation using CT-derived models, achieving sub-millimeter accuracy in phantom experiments. Later clinical studies, Ma~\emph{et al.}~\cite{ma2025augmented} and Altorfer~\emph{et al.}~\cite{altorfer2025pedicle}, confirmed that AR-assisted navigation can achieve accuracy comparable to robotic systems while improving workflow efficiency and minimizing fluoroscopy use. Beyond instrumentation, AR has also been applied to minimally invasive and percutaneous interventions. Agten~\emph{et al.}~\cite{agten2018augmented} reported improved targeting efficiency for lumbar facet joint injections, while Jiang~\emph{et al.}~\cite{jiang2023wearable} integrated ultrasound with wearable AR guidance to support lumbar puncture procedures.

Collectively, these advances in ultrasound imaging, robotic acquisition, and AR visualization illustrate a convergence toward adaptive, radiation-free image guidance for spinal interventions. However, most existing AR systems still rely on static preoperative CT or CBCT models that cannot adapt to intraoperative motion or deformation~\citep{fotouhi2019co,creighton2020early}. Integrating dynamic ultrasound-based reconstructions into AR environments could enable real-time, patient-specific visualization while maintaining spatial accuracy. Yet, no prior studies have systematically compared ultrasound- and CBCT-derived spine models within a unified AR-guided workflow—a gap that the present study aims to address through a quantitative and qualitative evaluation of both modalities for lumbar needle guidance.

In this study, we present a comparative evaluation of ultrasound-based and CBCT-based AR workflows for spinal needle insertion guidance. We integrate both imaging modalities into a unified AR framework, allowing direct spatial comparison and controlled testing of visualization accuracy, anatomical fidelity, and usability. Specifically, (1) ultrasound volumes are reconstructed via robotic scanning and shape completion algorithms, (2) CBCT volumes are segmented and rendered as preoperative models, and (3) both are co-registered to a shared coordinate system for AR overlay during simulated needle insertion tasks. Through quantitative and qualitative analyses, we aim to elucidate how these two imaging approaches complement each other and to identify conditions under which each modality offers distinct clinical advantages.
\section{Methods}\label{sec:Methods}

\begin{figure}
    \centering
    \includegraphics[width=0.85\linewidth]{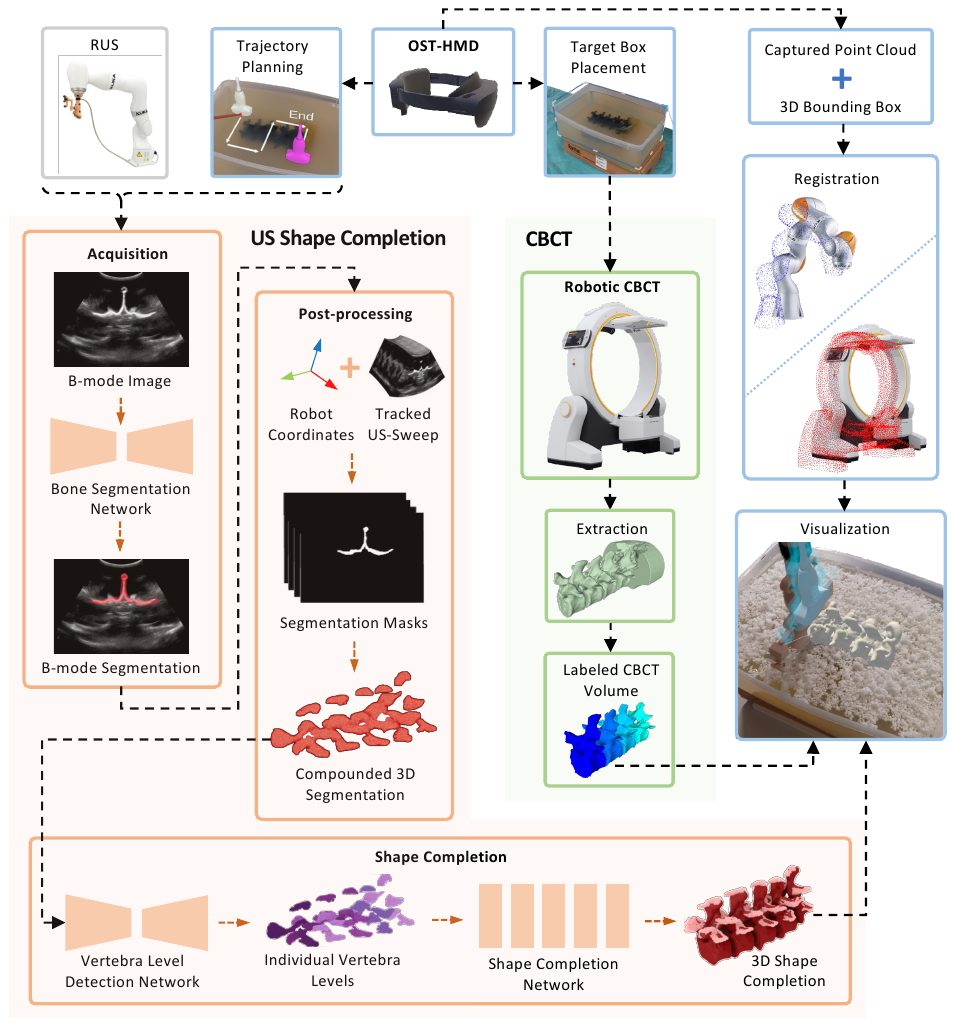}
    \caption{System Overview.}
    \label{fig:overview}
\end{figure}

Both pipelines follow a similar workflow: (1) planning the image acquisition in AR, (2) acquiring and reconstructing a 3D model of the lumbar spine, and (3) visualizing the model in AR to guide needle insertion. Figure~\ref{fig:overview} illustrates the overall system architecture. Both experiments were conducted on a phantom consisting of a 3D-printed spine model, a low-cost and reproducible design \cite{west2014development}, embedded in gelatin to simulate soft tissue and submerged in water. 

\subsection{Virtual-to-Real Registration}
The registration procedure establishes a shared coordinate system across both robotic devices and the optical see-through HMD.
The process begins by capturing 3D point clouds of the physical imaging device using the depth sensor integrated into the HMD. For the CBCT-based workflow, point clouds are captured of the CBCT machine gantry; for the ultrasound-based workflow, point clouds are captured of the robotic arm, as the CBCT machine is not present in this setup.
To isolate the device from the background environment, the user first specifies a 3D volume of interest. The system then gathers several depth captures from various perspectives, which are merged to form a comprehensive point cloud representing the real object
$R = \{\mathbf{r}_k \mid k = 1, \ldots, K\}$. A corresponding reference point cloud,
$M = \{\mathbf{m}_k \mid k = 1, \ldots, K\}$, is derived from the robot’s CAD model inside the AR environment. A two-step registration procedure aligns both point clouds. First, a global coarse alignment is computed using feature correspondences to estimate an initial rigid transformation $T$. The alignment is then refined using point-to-point iterative closest point (ICP) optimization:
\begin{equation}
    \hat{T} = \arg\min{T} \sum_k \|\,T\,\mathbf{s}_k - \mathbf{m}_{c(k)}\,\|^2
\end{equation}

where $c(k)$ denotes the correspondence mapping. The resulting transformation aligns the real device geometry with either the robotic ultrasound arm or the CBCT machine gantry in AR. This mapping allows any pose expressed in the AR frame to be transformed into the imaging device frame.

\subsection{Ultrasound-based Shape Completion Pipeline}

The pipeline leverages learned shape priors from high-resolution CT datasets to reconstruct complete vertebral geometries from incomplete ultrasound observations~\citep{gafencu2024shape}.
An AR interface was used to define the robotic ultrasound scanning trajectory directly in 3D space. Through an optical see-through HMD, the user visualized the virtual robot and spine anatomy in situ and placed virtual probe waypoints to specify the acquisition path. The planned trajectory was serialized and transmitted via a TCP/IP communication layer from the AR environment to the robotic control node, which executed the motion using Cartesian pose commands synchronized with the ultrasound acquisition. A zig-zag pattern was selected as the scanning protocol, as it produced denser and more uniformly distributed point clouds than linear or U-shaped paths, which improved mesh reconstruction and shape completion stability~\cite{gafencu2025shape}.

During robotic ultrasound acquisition, the system records tracked sweeps by synchronizing ultrasound image streams with robot end-effector poses. Each acquisition sequence generates a temporally coherent dataset where ultrasound frames are associated with corresponding transducer poses.
The shape completion algorithm processes these tracked sweeps to generate complete 3D vertebral models. The reconstruction process operates on segmented point clouds extracted from the compounded ultrasound volume, using a learned mapping function:

\begin{equation}
	P_{\text{complete}} = f_\theta (P_{\text{partial}}, C)
\end{equation}

where $P_{\text{partial}}$ represents the incomplete point cloud from ultrasound observations, $C$ encodes contextual shape priors, and $f_\theta$ is a neural network\footnote{Shape Completion Network: https://github.com/miruna20/Shape-Completion-in-the-Dark} trained on paired incomplete/complete vertebral geometries, based on the published work of Gafencu~\emph{et al.}~\cite{gafencu2024shape,gafencu2025shape}. 

The completed point clouds are converted to triangular meshes using Poisson surface reconstruction. To address bandwidth constraints for transmission to the HMD, the system employs Draco compression\footnote{Draco: https://google.github.io/draco/}. Mesh positioning within the AR scene is calculated based on the original scanning trajectory, ensuring accurate spatial correspondence between reconstructed anatomy and the physical scanning location.

\subsection{CBCT-based Visualization Pipeline}

Prior to image acquisition, the 3D scanning volume is defined interactively in AR through an optical see-through HMD. A 3D bounding box $\mathcal{B}$ representing the desired scan volume is positioned and scaled directly in 3D space to encompass the target lumbar levels. The bounding box is parameterized by its center position, orientation, and dimensions.
Once confirmed, these parameters are serialized into a transformation matrix $^{W}T_{\mathcal{B}}$ and transmitted via a TCP/IP protocol to the control workstation. On the imaging system, the received parameters are converted into the native CBCT coordinate frame $\mathcal{C}$ through the pre-calibrated transformation $^{C}T_{W}$. 
The final transformation describing the scan volume in the CBCT frame is therefore given by
\begin{equation}
^{C}\!T_{\mathcal{B}} = \,^{C}\!T_{W}\,^{W}\!T_{\mathcal{B}}.
\end{equation}
Transformations are applied from right to left, mapping coordinates from the Bounding Box in AR frame $\mathcal{B}$ to the World frame $W$, then to the CBCT frame $\mathcal{C}$. This mapping ensures that the volume planned in AR is aligned with the imaging space. 
The CBCT machine adjusts its source–detector geometry, gantry trajectory, and field of view, ensuring that the scan center and extent correspond to the planned volume.

Bone segmentation is performed in 3D Slicer\footnote{3D Slicer: https://www.slicer.org/} using intensity thresholding and morphological filtering to generate a binary mask, which is converted into a triangular surface mesh via the marching cubes algorithm.
The resulting mesh is downsampled using quadric edge decimation to balance real-time rendering efficiency and anatomical accuracy.

\section{Experiments}\label{Sec:UserStudy}

We conducted a between-subject user study to compare the two AR–guided imaging pipelines for spinal intervention. A total of 20 participants (6 female, 14 male; age range 22–43 years) were recruited from students and researchers with backgrounds in biomedical engineering or related medical imaging fields. Individuals with uncorrected visual acuity or uncorrected diplopia/strabismus that would interfere with the AR visualization were excluded to ensure consistency in visual perception. This was assessed via a self-report questionnaire prior to the study. In addition, participants self-reported their task-relevant experience on a 5-point Likert scale (1 = no familiarity, 5 = high familiarity). Overall, participants reported moderate familiarity with AR ($3.38 \pm 1.27$), ultrasound imaging ($3.42 \pm 1.20$), and robotic systems ($3.66 \pm 1.16$), indicating a baseline level of technical competence relevant to the evaluated workflows.
A between-subject design was chosen to avoid learning effects during the similar needle insertion phase between the workflows. Accordingly, the study is intended as a comparative evaluation of the two AR-guided workflows rather than an assessment of inter-individual performance differences or learning effects.
Each participant completed two phases: a planning and image acquisition phase, followed by a needle insertion phase. The goal was to evaluate how the imaging modality used in AR influences user performance, efficiency, and perceived usability.

The experimental setup consisted of a KUKA LBR iiwa 14 R820 robotic arm equipped with a Siemens ACUSON Juniper ultrasound machine (5C1 convex probe) and a Brainlab Loop-X robotic CBCT system. Both devices were spatially registered to the same AR coordinate frame, visualized through a Microsoft HoloLens 2 headset. A lumbar spine phantom was fixed in a water tank to provide acoustic coupling and realistic geometry for both scanning and insertion tasks. Participants interacted exclusively through the AR interface, which was connected to the robotic devices via a TCP/IP communication layer to synchronize model updates, probe or gantry positions, and rendered visualizations in real time.

In the planning and acquisition phase, participants used the AR interface to define the imaging region and initiate data acquisition. For ultrasound, this involved planning a zig-zag probe trajectory and launching an automated robotic scan; for CBCT, participants positioned a 3D bounding box to define the scan volume, which was then transmitted to the Loop-X system to automatically adjust its acquisition geometry. After imaging, the resulting 3D spinal model was displayed in AR. Planning time was recorded, and after each planning session, participants completed the NASA Task Load Index (NASA-TLX), System Usability Scale (SUS), and a 7-point Likert scale Single Ease Question (SEQ) questionnaires to assess workload, usability, and task difficulty.

In the needle insertion phase, participants used the same AR interface to plan and perform two needle insertion tasks on the phantom: a facet joint (FJ) injection and a lumbar puncture (LP). The target position was defined virtually in the CT volume in the CBCT workflow and in the 3D mesh of the ultrasound workflow. Afterwards, The target was transformed and visually indicated in the AR environment. Participants aligned the needle guide with the planned trajectory under AR visualization. Execution time and placement accuracy—defined as the Euclidean distance between the recorded needle tip and the pre-defined virtual target—were measured. To ensure consistent accuracy computation across modalities, a one-time spatial calibration between the robotic ultrasound system and the Loop-X CBCT was performed prior to the study, following the procedure described by Li~\emph{et al.}~\cite{li2025robotic}. This calibration established the fixed geometric relationship between the two imaging coordinate systems, enabling transformation of the needle tip position from the robot frame into the CBCT reference frame for quantitative evaluation. The calibration was used exclusively for post-experimental analysis and was not involved in any interactive AR visualization or guidance during the tasks. After completing both insertions, participants filled out the NASA-TLX, SUS, and SEQ questionnaires again, along with an additional adapted Trust in Automation questionnaire~\cite{jian2000foundations} to assess perceived reliability and confidence.

\section{Results}\label{Sec:Results}

All statistical analyses were performed using Python with SciPy~\footnote{Scipy: https://scipy.org/}. Data normality was first evaluated using the Shapiro–Wilk test. For measures that satisfied the normality assumption, two-tailed independent samples t-tests were used to determine statistical significance. When normality was violated, the Mann–Whitney U test was applied.

\subsection{Planning and Image Acquisition Phase}

\begin{figure}
    \centering
    \includegraphics[width=0.75\linewidth]{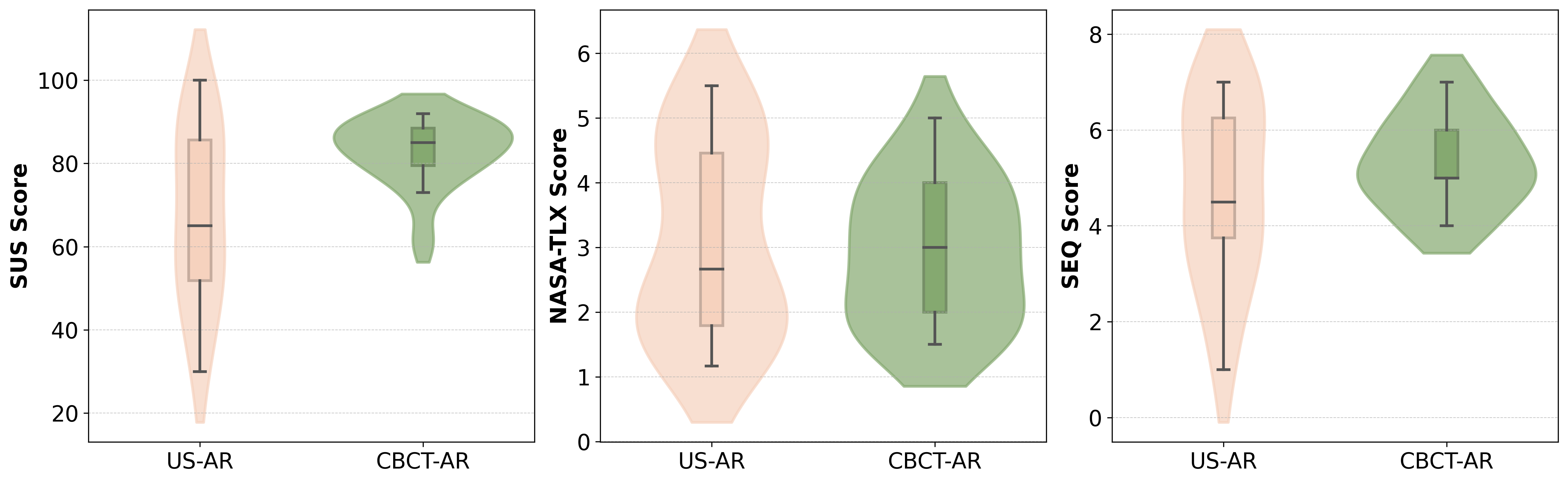}
    \caption{Subjective ratings for the planning phase comparing the ultrasound-based AR (US-AR) and CBCT-based AR (CBCT-AR) workflows.}
    \label{fig:planning}
\end{figure}

Planning time significantly differed between the two modalities, with participants using the CBCT-based AR system requiring on average $46.78 \pm 32.02$ s, compared to $83.50 \pm 34.42$ s in the ultrasound-based AR condition ($p < 0.001$). Subjective ratings for usability, workload, and task ease are shown in Figure \ref{fig:planning}. No statistically significant differences were found between the two modalities for subjective ratings during the planning phase. Task ease ($p = 0.299$), perceived workload ($p = 0.858$), and usability scores ($p = 0.069$) were comparable between conditions.
Overall, both AR systems were rated as usable and effective for scan planning.

For the CBCT-based AR guidance, the average radiation dose per scan was 24.84 mGy. This dose is consistent with clinical exposure levels reported for CBCT-guided lumbar spine procedure~\citep{berris2013radiation}. 

\subsection{Needle Insertion Phase}

Both groups successfully completed the two insertion tasks (FJ and LP) using their assigned imaging modality. The mean task times and placement errors are summarized in Table~\ref{tab:objective}. 
No statistically significant differences in insertion time were observed between the two workflows, either for individual tasks or combined time. For placement error, no significant difference was found for the facet joint task ($p = 0.3529$), while for the lumbar puncture, the CBCT-based workflow yielded significantly lower error ($p < 0.001$). The combined error across both tasks was also significantly lower for CBCT-based AR ($p < 0.001$).

\begin{table}
\centering
\caption{Needle insertion task completion time and error (mean$\pm$SD).}
\label{tab:objective}
\footnotesize
\setlength{\tabcolsep}{9pt}
\begin{tabular}{lccc}
\toprule
\textbf{Measure} & \textbf{US-AR Workflow} & \textbf{CBCT-AR Workflow} & \textbf{$p$ value} \\
\midrule
FJ time (s)            & 48.35$\pm$18.62 & 43.50$\pm$19.17 & 0.3327 \\
FJ error (mm)            & 11.28$\pm$3.70  & 9.69$\pm$2.93   & 0.3529 \\
LP time (s)            & 49.20$\pm$14.92 & 45.31$\pm$24.90 & 0.6212
 \\
LP error (mm)            & 24.33$\pm$2.69  & 8.89$\pm$2.71   & \textbf{\textless0.001} \\
\midrule
Combined time (s)      & 48.78$\pm$16.87 & 44.41$\pm$22.24 & 0.3138 \\
Combined error (mm)      & 17.81$\pm$7.28  & 9.29$\pm$2.85   & \textbf{\textless0.001} \\
\bottomrule
\end{tabular}
\end{table}

\begin{figure}
    \centering
    \includegraphics[width=0.95\linewidth]{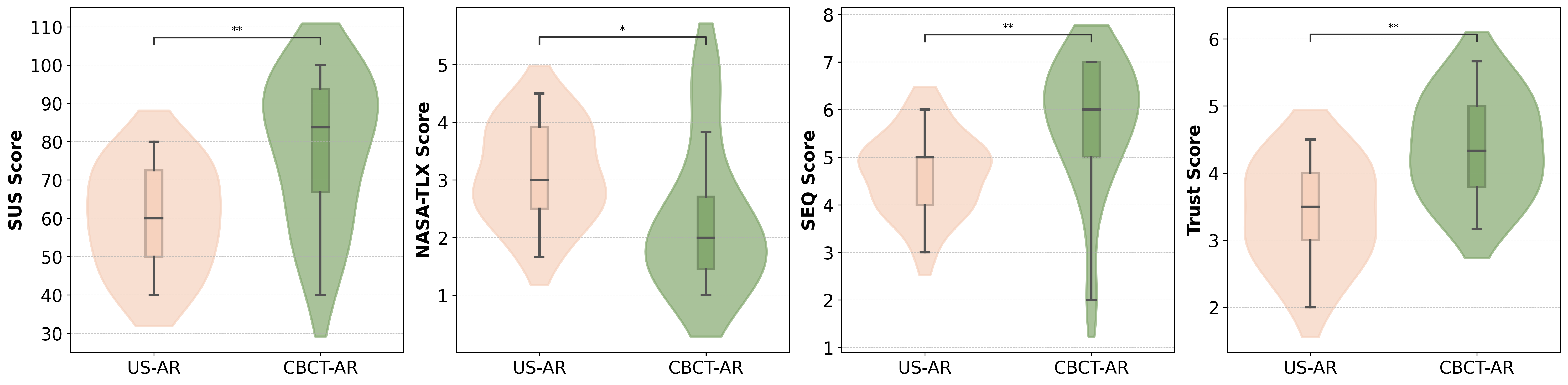}
    \caption{Subjective ratings for the needle insertion phase comparing the US-AR and CBCT-AR.}
    \label{fig:execution}
\end{figure}

Subjective ratings during the needle insertion phase are shown in Figure \ref{fig:execution}. Participants reported significantly higher usability ($p < 0.01$) and task ease ($p < 0.01$) with the CBCT-based AR system compared to the ultrasound-based condition. NASA-TLX scores indicated lower perceived workload for the CBCT-based system ($p < 0.05$) as well, suggesting that the higher quality visualization from CBCT reduced cognitive demand during alignment. Trust ratings were also significantly higher for CBCT-AR ($p < 0.01$), indicating that users perceived the CBCT-derived models as more reliable and stable for guidance than the ultrasound-based reconstructions.

\section{Discussion}\label{Sec:Discussion}

Overall, participants were able to complete both the planning and insertion phases successfully in either condition, demonstrating that AR guidance can effectively support procedural understanding and targeting in minimally invasive spine interventions.

Across both phases, participants using the CBCT-based AR system generally exhibited trends toward shorter task times and higher accuracy, with significantly better subjective ratings compared to those using the US-based system, particularly during the needle insertion phase. These findings are consistent with prior reports showing that high-resolution, preoperative imaging improves landmark clarity and spatial comprehension during AR navigation~\citep{liu2022spine, gibby2019head}. The CBCT-derived models provided continuous, complete visualization of the bony anatomy, allowing users to plan trajectories with greater confidence and minimal ambiguity regarding depth or orientation. The significantly higher trust and usability ratings observed for the CBCT condition further reflect users’ perception of reliability in the visualized anatomy.

In contrast, participants operating the ultrasound-based AR system experienced higher cognitive and temporal demand, as reflected in their NASA-TLX scores. This increased workload can be attributed to the intrinsic variability of ultrasound imaging, where acoustic shadowing and limited penetration through bone lead to incomplete local reconstructions. Although the probabilistic shape completion framework~\citep{gafencu2024shape, gafencu2025shape} effectively inferred vertebral geometry from partial ultrasound observations, it still relies on learned priors from generic datasets and may not perfectly capture patient-specific anatomical variations. The significantly larger placement error observed in the lumbar puncture task supports this interpretation: the target region was located in the lower lumbar spine, where the reconstructed model depended heavily on shape completion rather than direct ultrasound visibility. Such dependence on prior-based completion can reduce anatomical fidelity in regions that deviate from the training distribution or exhibit patient-specific variation. This observation highlights the challenge of generalizing learned priors to unseen anatomical configurations and underscores the need for adaptive refinement mechanisms that can update the completed surface using patient-specific observations.

The established clinical tolerance for safe and effective spinal needle placement is generally accepted to be $< 5$ mm~\citep{simpson2013use,simpson2022epidural}. While the mean absolute placement errors observed in this study (FJ: 9.69--11.28 mm, LP: 8.89--24.33 mm) exceed this threshold, it is important to distinguish between accuracy and precision. The low standard deviation across trials ($< 5$ mm) demonstrates high system precision and repeatability, suggesting that the systematic offset could potentially be corrected through improved calibration. This precision is critical for validating the technical reliability of the AR guidance framework, even as the absolute accuracy requires further refinement for clinical deployment.
Moreover, the observed facet joint targeting errors remained within the anatomical dimensions of the lumbar facets, typically 14–19 mm in width and 17–20 mm in height~\citep{gorniak2015lower}, indicating a safety margin context for the observed error for facet joint interventions.
In contrast, interspinous distances and spinous process dimensions are smaller (7–10 mm)~\citep{neumann1999determination}, meaning that errors of similar magnitude could compromise safety during lumbar puncture. This underscores the need for accurate reconstruction and trajectory planning for deeper or narrower anatomical targets.

Despite these differences in performance and trust, both systems achieved high overall usability, and the unified AR interface standardized the interaction experience across modalities. During the planning phase, usability and task ease were comparable, suggesting that the interface design and visualization layout minimized modality-related biases. The consistently high SUS scores observed in both conditions indicate that the AR-based visualization supports a usable and cognitively manageable interaction by presenting anatomical context directly within the user’s field of view, thereby limiting the need for frequent visual–motor reorientation~\citep{ma2025augmented, agten2018augmented}. However, as no baseline condition without AR guidance was included, the present study focuses on comparative AR-based workflows rather than on absolute changes in cognitive effort relative to conventional practice.

Several limitations should be acknowledged. First, all experiments were performed on a static phantom model, which does not reproduce the soft-tissue deformation, acoustic variability, and patient movement encountered in vivo. However, the study's primary goal was a controlled technical comparison of two imaging modalities within a unified AR framework.
Second, participants were recruited from students and researchers in biomedical engineering and medical imaging fields rather than practicing clinicians. These participants served as expert users to evaluate the technical and usability aspects of the novel AR interfaces, which is appropriate for a technical feasibility study at this stage of development. Clinical validation with interventional radiologists or spine surgeons will be essential in future work.
Additionally, the current setup required manual specification of scanning trajectories and needle paths, which was part of the workflow comparison to assess the user
interaction overhead of each modality. This could be improved through automated path optimization and integrated force or visual feedback to further streamline the workflow. Future work could focus on real-time multimodal fusion within a unified AR workflow to better emulate clinical conditions.
\section{Conclusion}\label{Sec:Conclusion}

This study presented a comparative evaluation of ultrasound-based and CBCT-based AR workflows for spinal needle interventions. Both modalities enabled intuitive in situ planning and guidance within the same AR framework. The CBCT-based AR system provided greater anatomical completeness, higher usability, and increased user trust, resulting in faster task completion and improved accuracy, particularly for deep or structurally complex targets. In contrast, the ultrasound-based system offered radiation-free, real-time adaptability but showed higher cognitive demand and reduced precision in regions reconstructed primarily through shape completion. These results suggest that combining static CBCT models for initial planning with dynamically updated ultrasound reconstructions could yield an adaptive, multimodal guidance strategy. Future work will focus on real-time multimodal fusion and in vivo validation to advance AR-assisted robotic imaging toward clinical integration.

\backmatter

\bibliography{09_Bibliography}

@article{liu2022spine,
  title={Spine surgery assisted by augmented reality: where have we been?},
  author={Liu, Yanting and Lee, Min-Gi and Kim, Jin-Sung},
  journal={Yonsei Medical Journal},
  volume={63},
  number={4},
  pages={305},
  year={2022}
}

@article{gibby2019head,
  title={Head-mounted display augmented reality to guide pedicle screw placement utilizing computed tomography},
  author={Gibby, Jacob T and Swenson, Samuel A and Cvetko, Steve and Rao, Raj and Javan, Ramin},
  journal={International journal of computer assisted radiology and surgery},
  volume={14},
  number={3},
  pages={525--535},
  year={2019},
  publisher={Springer}
}

@article{altorfer2025pedicle,
  title={Pedicle screw placement with augmented reality versus Robotic-assisted surgery},
  author={Altorfer, Franziska CS and Kelly, Michael J and Avrumova, Fedan and Burkhard, Marco D and Zhu, Jiaqi and Abel, Frederik and Cammisa, Frank P and Sama, Andrew and Farshad, Mazda and Lebl, Darren R},
  journal={Spine},
  volume={50},
  number={15},
  pages={1058--1064},
  year={2025},
  publisher={LWW}
}

@article{ma2025augmented,
  title={Augmented Reality Navigation System Enhances the Accuracy of Spinal Surgery Pedicle Screw Placement: A Randomized, Multicenter, Parallel-Controlled Clinical Trial},
  author={Ma, Yichao and Wu, Jiangpeng and Dong, Yanlong and Tang, Hongmei and Ma, Xiaojun},
  journal={Orthopaedic Surgery},
  volume={17},
  number={2},
  pages={631--643},
  year={2025},
  publisher={Wiley Online Library}
}

@article{agten2018augmented,
  title={Augmented reality--guided lumbar facet joint injections},
  author={Agten, Christoph A and Dennler, Cyrill and Rosskopf, Andrea B and Jaberg, Laurenz and Pfirrmann, Christian WA and Farshad, Mazda},
  journal={Investigative radiology},
  volume={53},
  number={8},
  pages={495--498},
  year={2018},
  publisher={LWW}
}

@article{jiang2023wearable,
  title={Wearable mechatronic ultrasound-integrated ar navigation system for lumbar puncture guidance},
  author={Jiang, Baichuan and Wang, Liam and Xu, Keshuai and Hossbach, Martin and Demir, Alican and Rajan, Purnima and Taylor, Russell H and Moghekar, Abhay and Foroughi, Pezhman and Kazanzides, Peter and others},
  journal={IEEE transactions on medical robotics and bionics},
  volume={5},
  number={4},
  pages={966--977},
  year={2023},
  publisher={IEEE}
}

@article{rekatsina2025importance,
  title={The Importance of Image Guidance in Common Spine Interventional Procedures for Pain Management: A Comprehensive Narrative Review},
  author={Rekatsina, Martina and Peng, Philip WH},
  journal={Pain and Therapy},
  pages={1--23},
  year={2025},
  publisher={Springer}
}

@article{kimura2023comparative,
  title={Comparative efficacy of ultrasound guidance and fluoroscopy or computed tomography guidance in spinal nerve injections: a systematic review and meta-analysis},
  author={Kimura, Ryota and Yamamoto, Norio and Watanabe, Jun and Ono, Yuichi and Hongo, Michio and Miyakoshi, Naohisa},
  journal={European Spine Journal},
  volume={32},
  number={12},
  pages={4101--4110},
  year={2023},
  publisher={Springer}
}

@article{viderman2023ultrasound,
  title={Ultrasound-guided vs. fluoroscopy-guided interventions for back pain management: a systematic review and meta-analysis of randomized controlled trials},
  author={Viderman, Dmitriy and Aubakirova, Mina and Aryngazin, Anuar and Yessimova, Dinara and Kaldybayev, Dastan and Tankacheyev, Ramil and Abdildin, Yerkin G},
  journal={Diagnostics},
  volume={13},
  number={22},
  pages={3474},
  year={2023},
  publisher={MDPI}
}

@article{gafencu2024shape,
  title={Shape completion in the dark: completing vertebrae morphology from 3D ultrasound},
  author={Gafencu, Miruna-Alexandra and Velikova, Yordanka and Saleh, Mahdi and Ungi, Tamas and Navab, Nassir and Wendler, Thomas and Azampour, Mohammad Farid},
  journal={International Journal of Computer Assisted Radiology and Surgery},
  volume={19},
  number={7},
  pages={1339--1347},
  year={2024},
  publisher={Springer}
}

@article{li20253d,
  title={3D ultrasound shape completion and anatomical feature detection for minimally invasive spine surgery},
  author={Li, Ruixuan and Cai, Yuyu and Davoodi, Ayoob and Borghesan, Gianni and Vander Poorten, Emmanuel},
  journal={Medical \& Biological Engineering \& Computing},
  pages={1--14},
  year={2025},
  publisher={Springer}
}

@article{gafencu2025shape,
  title={Shape Completion and Real-Time Visualization in Robotic Ultrasound Spine Acquisitions},
  author={Gafencu, Miruna-Alexandra and Shaban, Reem and Velikova, Yordanka and Azampour, Mohammad Farid and Navab, Nassir},
  journal={arXiv preprint arXiv:2508.08923},
  year={2025}
}

@article{azampour2024anatomy,
  title={Anatomy-aware computed tomography-to-ultrasound spine registration},
  author={Azampour, Mohammad Farid and Tirindelli, Maria and Lameski, Jane and Gafencu, Miruna and Tagliabue, Eleonora and Fatemizadeh, Emad and Hacihaliloglu, Ilker and Navab, Nassir},
  journal={Medical Physics},
  volume={51},
  number={3},
  pages={2044--2056},
  year={2024},
  publisher={Wiley Online Library}
}

@article{nagpal2015multi,
  title={A multi-vertebrae CT to US registration of the lumbar spine in clinical data},
  author={Nagpal, Simrin and Abolmaesumi, Purang and Rasoulian, Abtin and Hacihaliloglu, Ilker and Ungi, Tamas and Osborn, Jill and Lessoway, Victoria A and Rudan, John and Jaeger, Melanie and Rohling, Robert N and others},
  journal={International journal of computer assisted radiology and surgery},
  volume={10},
  number={9},
  pages={1371--1381},
  year={2015},
  publisher={Springer}
}

@article{navab2022medical,
  title={Medical augmented reality: definition, principle components, domain modeling, and design-development-validation process},
  author={Navab, Nassir and Martin-Gomez, Alejandro and Seibold, Matthias and Sommersperger, Michael and Song, Tianyu and Winkler, Alexander and Yu, Kevin and Eck, Ulrich},
  journal={Journal of Imaging},
  volume={9},
  number={1},
  pages={4},
  year={2022},
  publisher={MDPI}
}

@article{song2022happy,
  title={HAPPY: hip arthroscopy portal placement using augmented reality},
  author={Song, Tianyu and Sommersperger, Michael and Baran, The Anh and Seibold, Matthias and Navab, Nassir},
  journal={Journal of Imaging},
  volume={8},
  number={11},
  pages={302},
  year={2022},
  publisher={MDPI}
}

@article{fotouhi2020development,
  title={Development and pre-clinical analysis of spatiotemporal-aware augmented reality in orthopedic interventions},
  author={Fotouhi, Javad and Mehrfard, Arian and Song, Tianyu and Johnson, Alex and Osgood, Greg and Unberath, Mathias and Armand, Mehran and Navab, Nassir},
  journal={IEEE transactions on medical imaging},
  volume={40},
  number={2},
  pages={765--778},
  year={2020},
  publisher={IEEE}
}

@article{creighton2020early,
  title={Early feasibility studies of augmented reality navigation for lateral skull base surgery},
  author={Creighton, Francis X and Unberath, Mathias and Song, Tianyu and Zhao, Zhuokai and Armand, Mehran and Carey, John},
  journal={Otology \& Neurotology},
  volume={41},
  number={7},
  pages={883--888},
  year={2020},
  publisher={LWW}
}

@article{fotouhi2019co,
  title={Co-localized augmented human and X-ray observers in collaborative surgical ecosystem},
  author={Fotouhi, Javad and Unberath, Mathias and Song, Tianyu and Hajek, Jonas and Lee, Sing Chun and Bier, Bastian and Maier, Andreas and Osgood, Greg and Armand, Mehran and Navab, Nassir},
  journal={International journal of computer assisted radiology and surgery},
  volume={14},
  number={9},
  pages={1553--1563},
  year={2019},
  publisher={Springer}
}

@article{li2025robotic,
  title={Robotic CBCT meets robotic ultrasound},
  author={Li, Feng and Bi, Yuan and Huang, Dianye and Jiang, Zhongliang and Navab, Nassir},
  journal={International Journal of Computer Assisted Radiology and Surgery},
  pages={1--9},
  year={2025},
  publisher={Springer}
}

@article{berris2013radiation,
  title={Radiation dose from cone-beam CT in neuroradiology applications},
  author={Berris, Theocharis and Gupta, Rajiv and Rehani, Madan M},
  journal={American Journal of Roentgenology},
  volume={200},
  number={4},
  pages={755--761},
  year={2013},
  publisher={American Roentgen Ray Society}
}

@article{jian2000foundations,
  title={Foundations for an empirically determined scale of trust in automated systems},
  author={Jian, Jiun-Yin and Bisantz, Ann M and Drury, Colin G},
  journal={International journal of cognitive ergonomics},
  volume={4},
  number={1},
  pages={53--71},
  year={2000},
  publisher={Taylor \& Francis}
}

@article{gorniak2015lower,
  title={Lower lumbar facet joint complex anatomy},
  author={Gorniak, G and Conrad, W},
  journal={Austin J Anat},
  volume={2},
  number={1},
  pages={1--8},
  year={2015}
}

@article{neumann1999determination,
  title={Determination of inter-spinous process distance in the lumbar spine: evaluation of reference population to facilitate detection of severe trauma},
  author={Neumann, P and Wang, Y and K{\"a}rrholm, J and Malchau, H and Nordwall, A},
  journal={European Spine Journal},
  volume={8},
  number={4},
  pages={272--278},
  year={1999},
  publisher={Springer}
}

@article{simpson2022epidural,
  title={Epidural needle guidance using viscoelastic tissue response},
  author={Simpson, Benjamin Scott and Burns, Michael and Dick, Robert P and Saager, Leif},
  journal={IEEE Journal of Translational Engineering in Health and Medicine},
  volume={10},
  pages={1--11},
  year={2022},
  publisher={IEEE}
}

@article{simpson2013use,
  title={Use of ultrasound in chronic pain medicine. Part 1: neuraxial and sympathetic blocks},
  author={Simpson, Graham and Nicholls, Barry},
  journal={Continuing Education in Anaesthesia, Critical Care \& Pain},
  volume={13},
  number={5},
  pages={145--151},
  year={2013},
  publisher={Oxford University Press}
}

@article{west2014development,
  title={Development of an ultrasound phantom for spinal injections with 3-dimensional printing},
  author={West, Simeon J and Mari, Jean-Martial and Khan, Azalea and Wan, Jordan HY and Zhu, Wenjie and Koutsakos, Ioannis G and Rowe, Matthew and Kamming, Damon and Desjardins, Adrien E},
  journal={Regional Anesthesia \& Pain Medicine},
  volume={39},
  number={5},
  pages={429--433},
  year={2014},
  publisher={BMJ Publishing Group Ltd}
}

\end{document}